# Polylogarithmic Gap between Meshes with Reconfigurable Row/Column Buses and Meshes with Statically Partitioned Buses

Susumu Matsumae
Dept. of Information Science
Saga University, Japan

*Abstract*—This paper studies the difference in computational power between the mesh-connected parallel computers equipped with dynamically reconfigurable bus systems and those with static ones. The mesh with separable buses (MSB) is the mesh-connected parallel computer with dynamically reconfigurable row/column buses. The broadcast buses of the MSB can be dynamically sectioned into smaller bus segments by program control. We show that the MSB of size $n \times n$ can work with $O(\log^2 n)$ step even if its dynamic reconfigurable function is disabled. Here, we assume the word-model broadcast buses, and use the relation between the word-model bus and the bit-model bus.

## I. INTRODUCTION

The mesh-connected parallel computers equipped with dynamically reconfigurable bus systems gained much attention due to their strong computational powers [3, 11, 12, 13, 14]. The dynamic reconfigurable function enables the models to make efficient use of broadcast buses, and to solve many important, fundamental problems efficiently, mostly in a constant or polylogarithmic time [13]. Such reconfigurability, however, makes the bus systems complex and causes negative effects on the communication latency of global buses [2]. Hence, it is practically important to study the trade-off between such points quantitatively.

In this paper, we investigate the impact of reconfigurable capability on the computational power of mesh-connected computers with global buses. Here, we deal with the *meshes with separable buses* (MSB) [3, 12] and a variant of the meshes with partitioned buses called the *meshes with multiple partitioned buses* (MMPB) [4]. The MSB and the MMPB are the mesh-connected computers enhanced by the addition of broadcast buses along every row and column.

The broadcast buses of the MSB, called *separable buses*, can be dynamically sectioned into smaller bus segments by program control, while those of the MMPB, called *partitioned buses*, are statically partitioned in advance and cannot be dynamically reconfigurable. In the MSB model, each row/column has only one separable bus, while in the MMPB model, each row/column has $L$ partitioned buses ($L \geq 1$). By comparing the relative power between these models, we clarify the difference in computational power between the parallel models equipped with reconfigurable bus systems and those with static ones. In this paper, we assume that the size of MSB and that of MMPB are of $n \times n$. The case of different sizes was investigated in [8].

Here, we study how much slowdown is necessary when we deprive the MSB of its reconfigurable function. In [5, 6], we have shown that the MSB of size $n \times n$ can be simulated time-optimally in $O(n^{1/(2L+1)})$ steps using the MMPB of size $n \times n$, where $L$ is constant and the global buses are of word-model, i.e., the bus-width is the same as the number of bits in one word. From this result, it is natural to think that the slowdown may be at least of polynomial time. However, here we show that we can suppress the slowdown to polylogarithmic time, by making use of the relation between the word-model bus and the bit-model bus.

In this paper, we show that the $n \times n$ MSB can work with $O(\log^2 n)$ step slowdown even if its reconfigurable function is disabled. Here, we assume that the broadcast buses are of word-model, and use the relation between the word-model bus and the bit-model bus. As a corollary, since we have shown that the MSB of size $n \times n$ can simulate the reconfigurable mesh [1, 11, 14] (or PARBS, the processor array with reconfigurable bus systems) of size $n \times n$ in $O(\log^2 n)$ steps [10], we can say that the reconfigurable mesh of size $n \times n$ can also work with $O(\log^4 n)$ step slowdown even if its reconfigurable function is unused. In [7], we have proposed more efficient algorithm, which exploits the pipeline technique heavily. Although the algorithm presented here is slower than the one in [7] by the factor of $\log n$, the key ideas and explanations are much simpler than those in [7].

This paper is organized as follows: Section II describes the MSB and the MMPB models, and briefly explains how to solve the simulation problem of the MSB by using the MMPB. Section III shows that the $n \times n$ MSB can work with $O(\log^2 n)$ step slowdown even if its reconfigurable function is disabled. Lastly, Section IV offers concluding remarks.

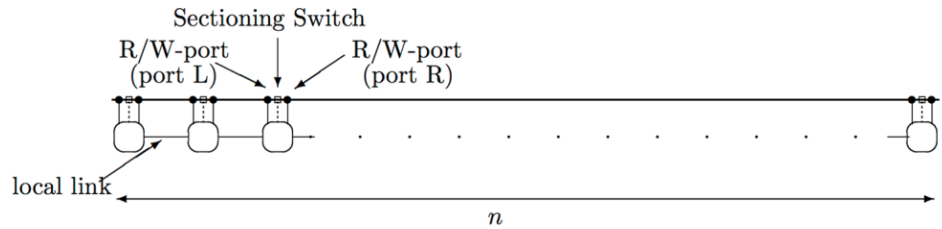

Figure 1. A separable bus along a row of the n × n MSB. Each PE has access to the bus via the two read/write-ports beside the sectioning switch.

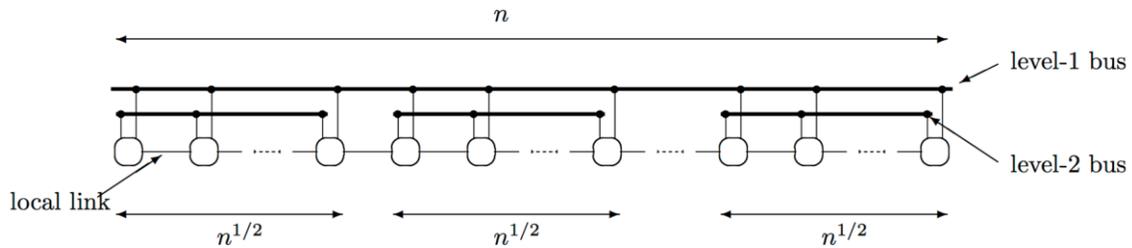

Figure 2. Partitioned buses along a row of the n × n MMPB. Here, L = 2, $\ell_1$ =n, and $\ell_2$ = $n^{1/2}$.

## II. PRELIMINARIES

### A. Models

An $n \times n$ mesh consists of $n$ identical processors or processing elements (PEs) arranged in a two-dimensional grid with $n$ rows and $n$ columns. We assume that all the meshes are synchronous. The PE located at the grid point $(i, j)$, denoted as PE$[i,j]$, is connected via bi-directional unit time communication links to those PEs at $(i \pm 1, j)$ and $(i, j \pm 1)$, provided they exist $(0 \leq i, j < n)$. PE[0,0] is located in the top-left corner of the mesh. Each PE$[i, j]$ is assumed to know its coordinates $(i, j)$.

An $n \times n$ mesh with separable buses (MSB) and an $n \times n$ mesh with multiple partitioned buses (MMPB) are the $n \times n$ meshes enhanced with the addition of broadcast buses along every row and column. The broadcast buses of the MSB, called separable buses, can be dynamically sectioned through the PE-controlled switches during the execution of programs, while those of the MMPB are statically partitioned in advance by a fixed length. In the MSB model, each row/column has only one separable bus (Fig. 1), while in the MMPB model each row/column has $L$ partitioned buses (Fig. 2). The MSB is essentially the same model as the horizontal-vertical reconfigurable mesh (HV-RM) described in [1, 13]. Those $L$ partitioned buses of the MMPB are indexed as level-1, level-2, ..., level-$L$, respectively. We assume that the partitioned buses of the MMPB are equally partitioned by the same length if they belong to the same level. For each level-$k$, the value $\ell_k$ denotes the length of a bus segment of the partitioned bus in level-$k$. Without loss of generality, we assume $\ell_k \geq \ell_{k+1}$.

We assume that the word size of processor is $\lceil \log n \rceil$ for a mesh of size $n \times n$. As for the bus-width, we consider two types of bus-models: word-model and bit-model [13]. In the word-model, a broadcast bus consists of $\lceil \log n \rceil$ wires and conveys one word of data in one step; in the bit-model, a broadcast bus consists of a single wire and conveys only one bit of data in a step. We call the MSB (resp. MMPB) with word-model global bus by the word-model MSB (resp. MMPB). The bit-model MSB and MMPB are termed similarly. (Strictly speaking, the bit-model defined in [13] assumes that both the processor word-size and bus-width are constant. Here, we assume that only bus-width is constant, and that processor word-size is of $\lceil \log n \rceil$ for the mesh of size $n \times n$.)

A single time step of the MSB and the MMPB is composed of the following three sub-steps:

1) *Local communication sub-step:*
   Every PE communicates with its adjacent PEs via local links.
2) *Broadcast sub-step:*
   Every PE changes its switch configurations by local decision (this operation is only for the MSB). Then, along each broadcast bus segment, several of the PEs connected to the bus send data to the bus, and several of the PEs on the bus receive the data transmitted on the bus.
3) *Compute sub-step:*
   Every PE executes some local computation.

Here, we assume that a PE writes to only one bus at a time in the MMPB model. The bus accessing capability is similar to that of the Common-CRCW PRAM model. If there is a write-conflict on a bus, the PEs on the bus receive a special value $\perp$ (i.e., PEs can detect whether there is a write-conflict on a bus or not). If there is no data transmitted on a bus, the PEs on the bus receive a special value $\phi$ (i.e., PEs can know whether there is data transmitted on a bus or not).

### B. Port-Connectivity-Graph

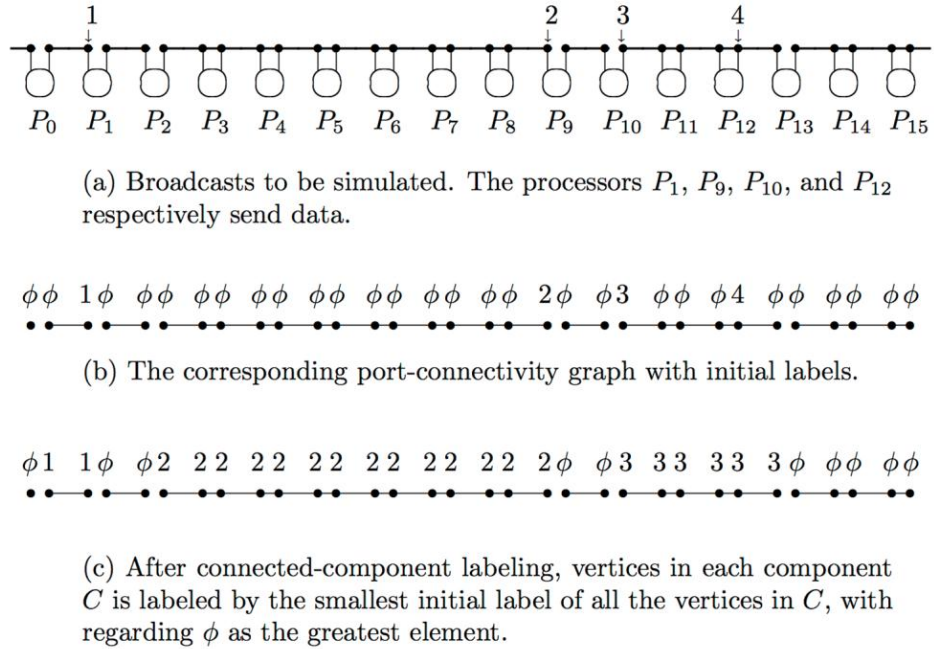

Figure 3. Broadcasts on a separable bus along a row of the $16 \times 16$ MSB are simulated by connected-component labeling of the port-connectivity graph.

To simulate operations of the MSB by using the MMPB, we focus on how to mimic the broadcast sub-step of the MSB by using the MMPB, because the local communication and the compute sub-steps of the MSB can be easily simulated in a constant number of steps by the MMPB.

The simulation of broadcast sub-step can be achieved by connected-component labeling (CC-labeling) of a *port-connectivity graph* (pc-graph). See Fig. 3 (a) and (b) for an example. Vertices of the pc-graph correspond to read/write-ports of PEs, and edges stand for the port-to-port connections. Each vertex is initially labeled by the value that is sent through the corresponding port by the PE at the broadcast sub-step. If there is no data sent through the port, the vertex is labeled by $\phi$. The CC-labeling is done in such a way that vertices in each component $C$ is labeled by the smallest initial label of all the vertices in $C$, with regarding $\phi$ as the greatest value (Fig. 3 (c)). These labels are called *component labels*. Obviously, the broadcast sub-step of the MSB can be simulated in $O(T)$ steps on the MMPB if the CC-labeling of the corresponding pc-graph can be solved in $O(T)$ steps by the MMPB. If there occurs collision on a bus segment, at least one of the senders can detect it by comparing its sending data with the component label obtained by the CC-labeling algorithm (e.g., $P_{12}$ in Fig. 3 senses the collision). Then, by distributing such collision information using the CC-labeling algorithm, PEs can resolve the collision.

In Section III, we solve the CC-labeling problem by the divide-and-conquer strategy composed of the following three phases:

**Phase 1:** *{local labeling}*

Divide the pc-graph into sub-graphs, and label vertices locally within each sub-graph. These labels are called *local component labels*. In each sub-graph, also check whether the two vertices located at the boundary of the sub-graph are connected to each other or not (this connectivity information is used in Phase 2).

**Phase 2:** *{global labeling of boundary vertices}*

Label those vertices located at the boundary of each sub-graph with component labels.

**Phase 3:** *{local labeling for adjustment}*

Update vertex labels with component labels within each of the sub-graph for the consistency with Phase 2.

In the next section, we implement the above algorithm on the MMPB model.

III. SIMULATION ALGORITHM

In this section, we show that the $n \times n$ MSB can work with $O(\log^2 n)$ step slowdown even if its reconfigurable function is disabled. For clarity, let MMPB$^{<L>}$ denote the MMPB that has $L$ partitioned buses per row/column.

First, we prove that any step of the word-model MSB of size $n \times n$ can be simulated by the word-model MMPB$^{<L>}$ of size $n \times n$ in $O(Ln^{1/(2L+1)})$ steps even when $L$ is non-constant. Next, we show that any step of the word-model MSB of size $n \times n$ can work with $O(\log^2 n)$ step slowdown even if we deprive the MSB of its reconfigurability, by considering the relation between the word-model bus and the bit-model one.

## A. Simulation of the word-model MSB by the word-model MMPB

In [5, 6], we have proved the following lemma, assuming that $L$ is a fixed constant.

**Lemma 1** [5, 6] *Any step of the word-model MSB of size $n \times n$ can be simulated by the word-model MMPB$^{<L>}$ of size $n \times n$ in $O(\sqrt{\ell_L} + \sum_{j=1}^{L-1} \sqrt{\ell_j/\ell_{j+1}} + n/\ell_1)$ steps where $L$ is a fixed constant.* ∎

In this section, we show that we obtain the almost same result as Lemma 1, even if we assume that $L$ is non-constant.

In what follows, we mainly focus on how to simulate the broadcasts along a row of the simulated MSB by using the corresponding row of the simulating MMPB. The simulation for columns can be achieved similarly.

To begin with, we introduce two fundamental results.

**Lemma 2** [9] *The broadcasts taken on the separable bus in the row $i$ of the word-model MSB of size $n \times n$ can be simulated in $O(\sqrt{\ell_1} + n/\ell_1)$ steps by the word-model MMPB$^{<1>}$ of size $n \times n$.* ∎

**Corollary 1** [9] *The broadcasts taken on the separable bus in the row $i$ of the word-model MSB of size $n \times n$ can be simulated in $O(\sqrt{n})$ steps by the word-model MMPB$^{<1>}$ of size $n \times n$ when $\ell_1 = n$.* ∎

Then, we can prove the following lemma even if $L$ is non-constant.

**Lemma 3** *The broadcasts taken on the separable bus in the row $i$ of the word-model MSB of size $n \times n$ can be simulated in the row $i$ of the word-model MMPB$^{<L>}$ of size $n \times n$ in $O(\sqrt{\ell_L} + \sum_{j=1}^{L-1} \sqrt{\ell_j/\ell_{j+1}} + n/\ell_1 + L)$ steps.*

Proof: Let define $T_k(n)$, $U(n)$, and $V(n)$ as follows:

$T_k(n)$: the time cost for simulating the broadcasts taken along the separable bus in row $i$ of the word-model MSB of size $n \times n$ using row $i$ of the word-model MMPB$^{<k>}$ of size $n \times n$.

$U(n)$: the time cost for simulating the broadcasts taken along the separable bus in row $i$ of the word-model MSB of size $n \times n$ by using row $i$ of the word-model MMPB$^{<1>}$ of size $n \times n$.

$V(n)$: the time cost for simulating the broadcasts taken along the row $i$ of the word-model MSB of size $n \times n$ by using row $i$ of the word-model MMPB$^{<1>}$ of size $n \times n$ when $\ell_1 = n$.

From Lemma 2 and Corollary 1, there exist some constants $c_1$ and $c_2$ such that the following two inequalities hold:

$$U(n) \leq c_1\left(\sqrt{\ell_1} + \frac{n}{\ell_1}\right), \qquad (1)$$

$$V(n) \leq c_2(\sqrt{n}). \qquad (2)$$

In what follows, we prove that the following equation holds for some constant $c$. The proof is done by mathematical induction on $k$ ($k \geq 1$).

$$T_k(n) \leq c\left(2\sqrt{\ell_k} + 2\sum_{j=1}^{k-1}\sqrt{\frac{\ell_j}{\ell_{j+1}}} + \frac{n}{\ell_1} + k\right) \qquad (3)$$

Here, without loss of generality, we assume $c \geq c_1, c_2$.

For the base case where $k = 1$, from Eq. (1) and $c \geq c_1$, we have

$$T_1(n) = U(n) \leq c_1\left(\sqrt{\ell_1} + \frac{n}{\ell_1}\right) \leq c\left(2\sqrt{\ell_1} + \frac{n}{\ell_1} + 1\right)$$

and thus Eq. (3) holds.

For the inductive case where $k > 1$, we prove Eq. (3), assuming that the following inductive hypothesis holds.

$$T_{k-1}(n) \leq c\left(2\sqrt{\ell_{k-1}} + 2\sum_{j=1}^{k-2}\sqrt{\frac{\ell_j}{\ell_{j+1}}} + \frac{n}{\ell_1} + k - 1\right) \qquad (4)$$

Let $P_j$ and $P'_j$ respectively denote PE[$i,j$] of the $n \times n$ MSB and PE[$i,j$] of the $n \times n$ MMPB$^{<k>}$ ($0 \leq j < n$). Now, we explain how to implement the algorithm defined in Section II B. We divide the pc-graph corresponding to the broadcasts on the row separable bus into $n/\ell_k$ disjoint sub-graphs $G_0, G_1, ..., G_{(n/\ell_1)-1}$ of width $\ell_k$. Here, we say that a sub-graph of pc-graph is of width $w$ if it contains $2w$ vertices corresponding to the read/write-ports of $w$ consecutive PEs. The CC-labeling of such defined pc-graph is carried out on the MMPB$^{<k>}$ as follows. We divide the row of the simulating MMPB$^{<k>}$ into $n/\ell_k$ disjoint blocks $B_0, B_1, ..., B_{(n/\ell_1)-1}$ in a way that each $B_p$ consists of $P'_j$ ($p\ell_k \leqq j < (p+1)\ell_k$). Note that each sub-graph $G_p$ is processed by block $B_p$ alone. Then, for each block $B_p$, since the PEs in $B_p$ and a bus segment of the level-$k$ partitioned bus can be seen as a linear processor array of $\ell_k$ PEs with a single broadcast bus of length $\ell_k$, Phase 1 can be executed in $V(\ell_k)$ steps. As for Phase 2, the number of active PEs is $2n/\ell_k$, and each of those PEs can communicate in a constant time with next such PEs via either a local communication link or a bus segment of the level-$k$ partitioned bus. Hence, by conveying the information of boundary vertices of each $G_p$ to the leftmost PE in $B_p$, and letting the information be processed by the leftmost PE in $B_p$ alone, Phase 2 is essentially the same problem as simulating the broadcast operation of the $1 \times n/\ell_k$ MSB using the $1 \times n/\ell_k$ MMPB$^{<k-1>}$ where each level-$j$ partitioned buses are segmented by the length $\ell'_j = \ell_j/\ell_k$ ($1 \leq j < k$). Here, It should be noted that $\ell_l \geqq \ell_{l+1}$ holds for each $l$ ($1 \leq l < k$). The operations required for such adjustment (data transmission to/from the leftmost PE of each $B_p$, etc.) can be completed in a constant number of steps, and let $c_3$ be the time cost for them. Without violating the argument here, we assume that $c \geq c_3$ holds. (We can chose the constant $c$ appropriately in advance so that $c \geqq c_3$ holds.) Then, from Eq. (4), Phase 2 can be completed in

$$T_{k-1}\left(\frac{n}{\ell_k}\right) + c_3$$

where level-$j$ bus is segmented by $\ell'_j = \ell_j/\ell_k$

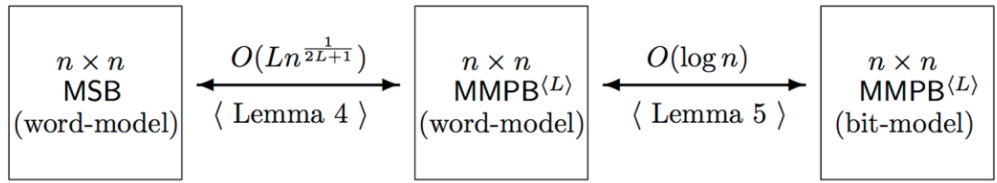

Figure 4. The simulation costs among the MSB and MMPBs.

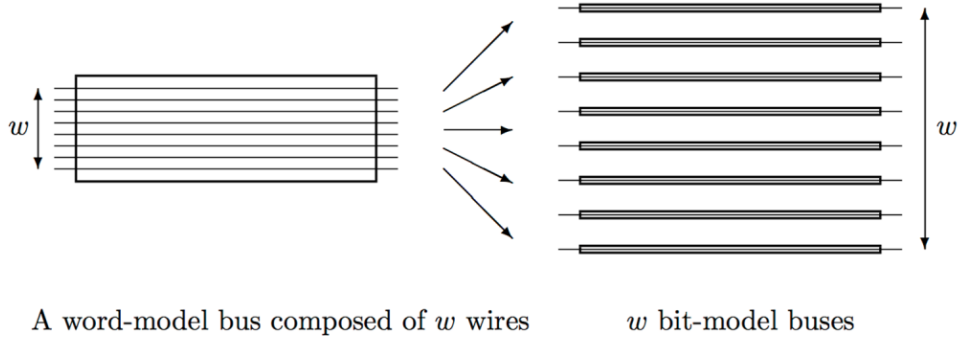

Figure 5. A word-model bus is decomposed to w bit-model buses where w is the word-length of processor.

$$\leq \quad \langle \text{Eq.}(4), c \geq c_3 \rangle$$

$$c\left(2\sqrt{\ell'_{k-1}} + 2\sum_{j=1}^{k-2}\sqrt{\frac{\ell'_j}{\ell'_{j+1}}} + \frac{n}{\ell_k}/\ell'_1 + k\right)$$

$$=$$

$$c\left(2\sqrt{\frac{\ell_{k-1}}{\ell_k}} + 2\sum_{j=1}^{k-2}\sqrt{\frac{\ell_j}{\ell_k}\bigg/\frac{\ell_{j+1}}{\ell_k}} + \frac{n}{\ell_k}\bigg/\frac{\ell_1}{\ell_k} + k\right)$$

$$=$$

$$c\left(2\sum_{j=1}^{k-1}\sqrt{\frac{\ell_j}{\ell_{j+1}}} + \frac{n}{\ell_1} + k\right)$$

steps. Phase 3 can be done in $V(\ell_k)$ steps similarly to Phase 1. As a whole, the algorithm can be executed in

$$2V(\ell_k) + c\left(2\sum_{j=1}^{k-1}\sqrt{\frac{\ell_j}{\ell_{j+1}}} + \frac{n}{\ell_1} + k\right)$$

$$\leq \quad \langle \text{Eq.}(2) \rangle$$

$$2c_2\sqrt{\ell_k} + c\left(2\sum_{j=1}^{k-1}\sqrt{\frac{\ell_j}{\ell_{j+1}}} + \frac{n}{\ell_1} + k\right)$$

$$\leq \quad \langle c \geq c_2 \rangle$$

$$c\left(2\sqrt{\ell_k} + 2\sum_{j=1}^{k-1}\sqrt{\frac{\ell_j}{\ell_{j+1}}} + \frac{n}{\ell_1} + k\right)$$

steps, and thus Eq. (3) holds for $k >1$ as well. The conclusion follows. ∎

The local communication and the compute sub-steps of the MSB can be easily simulated in a constant number of steps by the MMPB. Hence, by letting each $\ell_j = \Theta(n^{\alpha_j})$ where $\alpha_j = 2(L - j + 1)/(2L + 1)$, we have the following lemma from Lemma 3.

**Lemma 4** *Any step of the word-model MSB of size $n \times n$ can be simulated by the word-model MMPB$^{<L>}$ of size $n \times n$ in $O(Ln^{1/(2L+1)})$ steps.* ∎

### B. Simulation of the word-model MSB by the bit-model MMPB

In this section, we show that any step of the word-model MSB of size $n \times n$ can work with $O(\log^2 n)$ step slowdown even if we deprive the MSB of its reconfigurable function, by considering the relation between the word-model bus and the bit-model one.

First, we prove the following lemma.

**Lemma 5** *Any step of the word-model MMPB$^{<L>}$ of size $n \times n$ can be simulated by the bit-model MMPB$^{<L>}$ of size $n \times n$ in $O(\log n)$ steps.*

Proof: The $\lceil \log n \rceil$ bits of one word data can be conveyed sequentially in $\lceil \log n \rceil$ steps, one bit per step, in the bit-model MMPB$^{<L>}$. ∎

We illustrate the results of Lemma 4 and 5 in Fig. 4. Obviously, Fig. 4 implies the following lemma:

**Lemma 6** *Any step of the word-model MSB of size $n \times n$ can be simulated by the bit-model MMPB$^{<L>}$ of size $n \times n$ in $O(Ln^{1/(2L+1)} \log n)$ steps.* ∎

By letting $L = \log n$, we obtain the following corollary.

**Corollary 2** *Any step of the word-model MSB of size $n \times n$ can be simulated by the bit-model MMPB$^{<\log n>}$ of size $n \times n$ in $O(log^2 \, n)$ steps.*

Proof: By letting $L = \log n$, we can calculate the time-complexity as follows:

$$O(Ln^{1/(2L+1)}\log n)$$

$= \langle \; L = \log n \; \rangle$

$$O(n^{1/(2\log n+1)}\log^2 n)$$

$= \langle \; n^{1/(2\log n+1)} \leq c \; \text{ for some constant } c \; \rangle$

$$O(\log^2 n)$$

Thus, the conclusion follows. ∎

Since the word-model MSB of size $n \times n$ has $\log n$ wires for each row/column, we can view a word-model bus as $\log n$ bit-model buses (Fig. 5). Hence, without increasing any circuit-complexity, we obtain the bit-model MMPB$^{<\log n>}$ of size $n \times n$ from the word-model MSB of size $n \times n$. With this observation and Corollary 2, we can state the main theorem of this paper as follows:

**Theorem 1** *Any step of the word-model MSB of size $n \times n$ can work with $O(\log^2 n)$ step slowdown even if its reconfigurable capability is unused.* ∎

## IV. CONCLUDING REMARKS

In this paper, we showed that the word-model MSB of size $n \times n$ can work with $O(\log^2 n)$ step slowdown even if its reconfigurable capability is unused. We obtain the result from these two facts: 1) every global bus of the word-model MSB of size $n \times n$ consists of $\lceil \log n \rceil$ wires, and 2) we can obtain the bit-model MMPB of size $n \times n$ with $L=\lceil \log n \rceil$ from the word-model MSB of size $n \times n$ without increasing circuit-complexity. In [7], we have proposed more efficeint algorithm that exploits the pipeline technique heavily. Although the simulation algorithm presented here is slower than the one in [7] by the factor of $\log n$, the key ideas and explanations are much simpler than those in [7].

From a practical viewpoint, we expect that the communication latency of the broadcast buses of the MMPB is much smaller than that of the MSB. Each broadcast bus of the MSB of size $n \times n$ can form the broadcast bus whose length is $n$, and such a bus contains $O(n)$ sectioning switch elements in it. As for the MMPB of size $n \times n$, though the bus length is also at most $n$, but no switch element is inserted to the bus because it has no sectioning switch. Hence, compared to the MSB, the MMPB model has an advantage that each broadcast bus has smaller propagation delay introduced by the switch elements inserted into the bus (i.e., device propagation delay), and thus our simulation algorithm is practically useful when the mesh size becomes so large that we cannot neglect the delay. In future work, we will study the effectiveness of our simulation algorithm, by taking into account the propagation delay.

REFERENCES

[1] Y. Ben-Asher, D. Gordon, A. Schuster, "Efficient self-simulation algorithms for reconfigurable arrays," J. of Parallel and Distributed Computing 30 (1), 1−22, 1995

[2] T. Maeba, M. Sugaya, S. Tatsumi, K. Abe, "An influence of propagation delays on the computing performance in a processor array with separable buses," IEICE Trans. A J78-A (4), 523−526, 1995

[3] T. Maeba, S. Tatsumi, M. Sugaya, "Algorithms for finding maximum and selecting median on a processor array with separable global buses," IEICE Trans. A J72-A (6), 950−958, 1989

[4] S. Matsumae, "Simulation of meshes with separable buses by meshes with multiple partitioned buses," Proc. of the 17th International Parallel and Distributed Processing Symposium (IPDPS2003), IEEE CS press, 2003.

[5] S. Matsumae, "Optimal simulation of meshes with dynamically separable buses by meshes with statically partitioned buses," Proc. of the 2004 International Symposium on Parallel Architectures, algorithms and Networks (I-SPAN'04), IEEE CS Press, 2004.

[6] S. Matsumae, "Tight bounds on the simulation of meshes with dynamically reconfigurable row/column buses by meshes with statically par-titioned buses," J. of Parallel and Distributed Computing 66 (10), 1338−1346, 2006

[7] S. Matsumae, "Impact of reconfigurable function on meshes with row/column buses," International Journal of Networking and Computing 1 (1), 36−48, 2011

[8] S. Matsumae, "Effective Partitioning of Static Global Buses for Small Processor Arrays," Journal of Information Processing Systems 7 (1), 2011

[9] S. Matsumae, N. Tokura, "Simulating a mesh with separable buses," Transactions of Information Processing Society of Japan 40 (10), 3706−3714, 1999

[10] S. Matsumae, N. Tokura, "Simulation algorithms among enhanced mesh models," IEICE Transactions on Information and Systems E82−D (10), 1324−1337, 1999

[11] R. Miller, V. K. Prasanna-Kumar, D. Reisis, Q. F. Stout, "Meshes with reconfigurable buses," Proc. of the 5th MIT Conference on Advanced Research in VLSI, Boston, 1988.

[12] M. J. Serrano, B. Parhami, "Optimal architectures andalgorithms for mesh-connected parallel computers with separable row/column buses," IEEE Trans. Parallel and Distributed Systems 4 (10), 1073−1080, 1993

[13] R. Vaidyanathan, J. L. Trahan, Dynamic Reconfiguration, Kluwer Academic/Plenum Publishers, 2004.

[14] B. Wang, G. Chen, "Constant time algorithms for the transitive closure and some related graph problems on processor arrays with reconfigurable bus systems," IEEE Trans. Parallel and Distributed Systems 1 (4), 500−507, 1990